\documentclass[10pt,letterpaper,twocolumn]{article} 

\usepackage{ol2}
\usepackage{amsmath}

\begin{document}

\twocolumn[ 

\title{A microfabricated surface ion trap on a high-finesse optical mirror}


\author{Peter F. Herskind,$^{*}$ Shannon X. Wang, Molu Shi, Yufei Ge, Marko Cetina, and Isaac L. Chuang}

\address{MIT-Harvard Center for Ultracold Atoms, Department of Physics, Massachusetts Institute of Technology, Cambridge, Massachusetts 02139, USA \\
$^*$Corresponding author: herskind@mit.edu
}

\bibliographystyle{osajnl}

\begin{abstract}
A novel approach to optics integration in ion traps is demonstrated based on a surface electrode ion trap that is microfabricated on top of a dielectric mirror. Additional optical losses due to fabrication  are found to be as low as 80~ppm for light at 422~nm. The integrated mirror is used to demonstrate light collection from, and imaging of, a single $^{88}$Sr$^+$ ion trapped $169\pm4~\mu$m above the mirror. 
\end{abstract}


 ] 
Integration of optics with single atomic particles is of considerable interest in the exploration of the basic quantum physics of atom-light interactions as well as for the advancement of quantum information science and cavity quantum-electrodynamics (CQED). Much progress has been made in integrating mirrors, lenses, and optical fibers within hundreds of micrometers from neutral atoms\cite{Purdy2008,Wilzbach2009} and trapped ions\cite{VanDevender2010,Streed2011,Brady2011,Wilson2011,Merrill2011}. High finesse optical cavities around ions have also been investigated, but thus far, only with relatively large, millimeter-scale distances between ions and mirror surfaces~\cite{Mundt2002,Herskind2009a,Leibrandt2009a}. Integration of smaller cavities is challenged by the fact that motional states of trapped ions have been observed to decohere anomalously rapid, as $1/d^4$, for ion-surface distances $d$~\cite{Epstein2007}, and also by the observation that light on dielectric surfaces near trapped ions can cause charging and disruption of trap equilibrium~\cite{Harlander2010}. Much closer positioning of ions to mirrors is desirable in e.g. CQED systems~\cite{Kimble1998} where the ion-cavity couping strengths are inversely proportional to the cavity length, which scales approximately as $d$.  Moreover, accurate positioning of the ion relative to the cavity mode is essential, and it is clear that traditional bulk assembly techniques will not scale well to the systems envisioned for trapped ion quantum computation~\cite{Kim2005}.

\begin{figure}
\includegraphics[width=1\columnwidth]{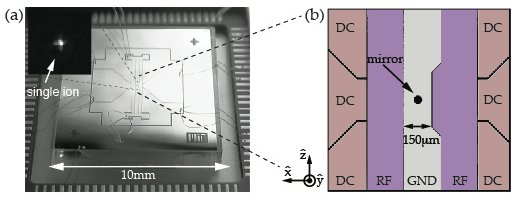}%
\caption{(a) Microfabricated ion trap on high reflectivity mirror. Inset shows a single trapped ion. (b) False color microscope image of the central region of the trap with all electrodes labelled. The ion is trapped $169~\mu$m above the central mirror aperture (black circle).
\label{fig:Setup}}%
\end{figure}

Here, we report on the demonstration of direct and scalable integration of an ion trap with a high reflectivity mirror, through microfabrication of a surface electrode ion trap~\cite{Chiaverini2005a} on the mirror. A circular aperture in the central electrode, located $169~\mu$m underneath the trap center (Fig.~\ref{fig:Setup}), allows the mirror to collect fluorescence and to image a single atomic ion. Despite its proximity, the presence of the mirror does not significantly perturb the trap, which is supported by the observation that trapping is stable with laser cooled ion lifetimes of several hours and with minimal sensitivity to light-induced charging. Furthermore, operation of the trap at 15~K aides to suppress anomalous ion heating~\cite{Labaziewicz2008}. This approach to integration of mirrors, and optics in general, in ion traps is scalable to large number of ion traps as multiple trapping zones with mirror apertures, may be defined on the same substrate with no additional overhead for fabrication.

The ion trap is fabricated on a 1.6~mm thick fused silica substrate that has a highly reflective dielectric coating optimized for light near 422~nm. The coating is composed of alternating layers of Ta$_2$O$_5$ and SiO$_2$ for a total thickness of $2.2~\mu$m and has been deposited by Advanced Thin Films using ion beam sputtering. Prior to fabrication, the substrate is cleaned by mechanical rubbing with cotton swabs and lens tissue using acetone, methanol and isopropanol, sequentially. Following a 5 min pre-bake at $110^\circ$C, the substrate is coated with NR9-3000PY photoresist spun at 3000~rpm and subsequently baked again at $110^\circ
$C for 5 min. Lithography is carried out using a chrome mask exposed at $3300~\mu$W/cm$^2$ for 2~min followed by a bake at $110^\circ$C for 2~min. The exposed traps are then developed in RD6 for 17~s, followed by a rinse in deionized water. Electrodes consisting of 10~nm Ti adhesion layer and a 400~nm layer of Ag are deposited by ebeam evaporation at a rate of 5~\AA/s. Finally the subtrate is soaked in acetone ($>99.9\%$) for 25~min until lift-off is completed, and then rinsed in methanol ($>99.9\%$). Figure~\ref{fig:Setup}(a) shows a picture of the finished trap, mounted in a ceramic pin grid array.

\begin{figure}
\includegraphics[width=1\columnwidth]{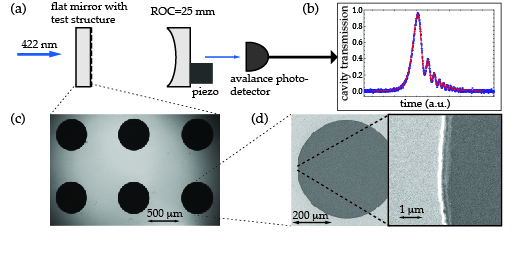}%
\caption{(a) Schematic of setup used for in loss measurements. (b) Example ring-down measurement by fast piezo scan. Fit is based on the model of Ref.~\cite{Poirson1997}. (c) Microscope image of test structure with array of mirror apertures. (d) SEM image of a test mirror aperture, following sputtering of $4$~nm layer of Au. Rightmost image shows detailed view of the edge. \label{fig:MirrorLoss}}%
\end{figure}

The mirror quality, prior to fabrication, is evaluated using ring-down spectroscopy~\cite{Poirson1997} in a near-confocal Fabry-Perot cavity setup. We find the cavity losses to be in agreement with the vendor specifications of a 45~ppm transmission coefficient and scattering and absorption losses of 25~ppm. To determine the losses incurred by the fabrication process, a test structure is fabricated in parallel with the trap using the same recipe. This structure effectively creates an array of test mirrors for which losses are evaluated using ring-down spectroscopy in a near-confocal Fabry-Perot cavity setup as shown in Fig.~\ref{fig:MirrorLoss}~(a) and (b). The second cavity mirror, on which no fabrication is done, has ROC=25~mm, resulting in a cavity mode waist on the $500~\mu$m diameter test mirrors [Fig.~\ref{fig:MirrorLoss}~(c)] of $40~\mu$m, which is sufficiently small that clipping losses on the apertures are negligible. Fig.~\ref{fig:MirrorLoss} (d) shows an SEM image of the mirror aperture, revealing clean edges with no visible residue. For a total of 15 test mirrors, we find an average increase in losses of $130\pm 10$~ppm and in the best cases the increase is at the level of 80~ppm.

We test the trap in a helium bath cryostat operated at 15~K. The trap has been described previously in~\cite{Labaziewicz2008} with the only difference that the trap design used in this work includes a $50~\mu$m diameter aperture in the central ground electrode directly below the ion. $^{88}$Sr$^+$ ions are loaded by resonant photoionization of a thermal vapor and subsequently Doppler laser cooled on the 5$^2$S$_{1/2}$$\leftrightarrow$5$^2$P$_{1/2}$ transition with light at 422~nm, while driving the 4$^2$D$_{3/2}$$\leftrightarrow$5$^2$P$_{1/2}$ repumping transition at 1092~nm. Upon loading of ions in the trap, we observe that these are trapped stably for hours with Doopler cooling, essentially only limited by the liquid helium hold time of the cryostat. 

\begin{figure}
\includegraphics[width=1\columnwidth]{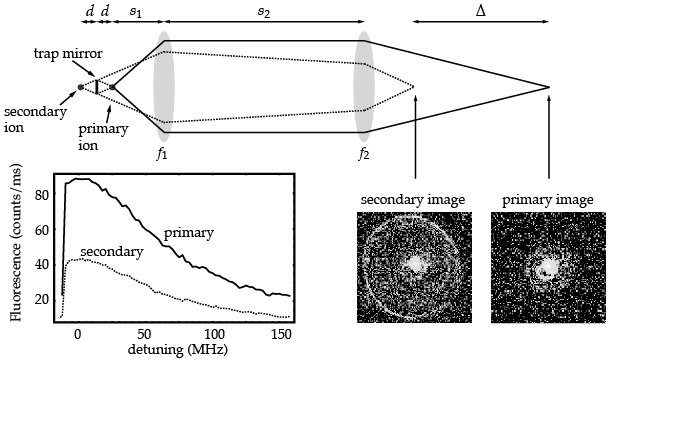}%
\caption{Schematic of imaging system: $f_1=25$~mm aspheric lens, $f_2=200$~mm acromat lens. Images show an ion imaged via the direct path through imaging system (primary) and via the mirror embedded in the trap (secondary). Graph shows the corresponding number of photons detected when scanning the 422~nm Doppler cooling laser across atomic resonance. The ratio of counts does not reflect the exact ratio of numerical apertures for the imaging paths due to imperfect spatial filtering.\label{fig:ImagingSystem}}%
\end{figure}

The ions can be imaged, by collecting scattered 422~nm light from the excited 5$^2$P$_{1/2}$ state, either directly, with an $\mathrm{NA}\sim0.45$ lens mounted inside the chamber, or via the integrated mirror, which subtends a solid angle corresponding to $\mathrm{NA}\sim0.15$ (Fig.~\ref{fig:ImagingSystem}). In both cases the light is directed onto a photomultiplier tube and a CCD camera outside the vacuum system.Two distinct images are thus formed~\cite{Allcock2010} and by appropriate adjustment of the lenses in the imaging system either image can be observed, as shown in Fig.~\ref{fig:ImagingSystem}. For our experimental configuration, where $d\ll f_1$ and $s_1-f_1\ll 1$, the ion height $d$ is related to the relative displacement of the imaging plane $\Delta$ as $d\simeq\frac{1}{2}\Delta (f_1/f_2)^2$. We measure $\Delta$ to $21\pm1~$mm, corresponding to an ion height $d=169\pm4~\mu$m in good agreement with numerical predictions of $165~\mu$m from boundary element analysis of the trap. This determination of $d$ is limited by the depth of focus of our imaging system but with ideal optics this method could in principle operate at the diffraction limit. 

While the geometry of our present design does not allow detection of ion fluorescence with efficiencies beyond those of standard bulk optics, installed separate from the trap, it demonstrates a basic concept by which light collection optics may be integrated into microfabricated surface electrode ion traps in a scalable fashion. As an example, Fig.~\ref{fig:ImagingSystem} shows a measurement of the 5$^2$S$_{1/2}$$\leftrightarrow$5$^2$P$_{1/2}$ transition of a single $^{88}$Sr$^+$ ion, where fluorescence is collected by adjusting the imaging system to either the primary or the secondary image of the ion, thus demonstrating a spectroscopic measurement using the integrated ion-mirror system. 

Sensitivity to laser-induced charging~\cite{Harlander2010}, which can potentially impose severe limitations to the practicality of experiments due to the proximity of the lasers to both trap and mirror, can be studied by deliberately exposing the trap to excess laser light and monitoring the effect on the ion. Tests are performed with light at 405~nm, 461~nm and 674~nm where about $200~\mu$W of power focused to a $\sim50~\mu$m radius spot is incident at grazing angle across the trap surface under the ion to simulate the effect of misaligned laser beams. The ion displacement, as a result of charge build-up, can be measured via the induced micromotion as the ion is displaced from the node of the rf-field~\cite{Berkeland1998} and quantified in terms of the adjustment of the trap voltages [DC in Fig.~\ref{fig:Setup}(a)] required to compensate this effect. Following this procedure, we observe only low sensitivity to charging for the wavelengths studied and the required changes in the DC voltages after 10~min of continuous exposure are at the level of 5~mV to 50~mV. The strongest effect is observed with light at 405~nm and corresponds to an induced field at the ion location of about 20~V/m. With no excess laser light incident on the electrodes and mirror, trapping is observed to be stable without the need for adjustment of DC voltages over a time span of an hour. 

Anomalous heating in ion traps is known to significantly increase with ion-trap distance~\cite{Epstein2007} and is evaluated here as described in~\cite{Labaziewicz2008}. The measurements are done at a secular frequency of $\omega_z=2\pi\times0.7$~MHz and the lowest heating rate observed is $0.10\pm0.01$~quanta/ms. A spectral density $S$ of the fluctuating fields driving the heating can be deduced from the heating rate~\cite{Epstein2007} and expressed as $\omega_z S(\omega_z)=(3.6\pm0.4)\times10^{-6}$~V$^2/$m$^2$, which is comparable to or lower than traps of similar dimensions operated at room temperature~\cite{Epstein2007}. The result is about an order of magnitude higher than the lowest heating rate obtained previously by our group for a cryogenic trap of the same electrode material and geometry but without the aperture and the dielectric mirror coating~\cite{Labaziewicz2008}. Non-contact friction measurements using cantilevers have similarly observed about an order of magnitude increase for a bare fused silica substrate relative to a gold surface~\cite{Stipe2001}; however, further investigation is required to determine if the heating rate observed here is influenced by the exposed dielectric. 

In conclusion, we have demonstrated a novel concept for integration of optics and ion traps. Our microfabrication procedure does not compromise the mirror quality, significantly, and ion trapping is not adversely affected by the presence of the exposed dielectrics. The collection of ion fluorescence complements recent ion trap experiments with integrated multi-mode optical fibers and phase Fresnel lenses~\cite{VanDevender2010,Streed2011,Brady2011,Wilson2011,Merrill2011}. We note that the limited numerical aperture of our present system is not fundamental and can be increased by appropriate changes to the trap geometry and can also be optimized to collimate the light, e.g. by laser machining~\cite{Hunger2010a} or chemical etching~\cite{Merrill2011,Noek2010} of concave mirrors into the trap substrate. Furthermore, our approach circumvents issues pertaining to the alignment of the mirror relative to the trap, is easily scalable to large numbers of ion-mirror systems and is compatible with single mode optics.

Our design concept furthermore provides a pathway for construction of ion-cavity systems in the framework of CQED, where a complete cavity-ion system is achieved by adding a second concave mirror above the trap. The low ion height allows for a sub-millimeter cavity length and can thus potentially achieve the low mode volume required to reach the strong coupling regime of CQED~\cite{Kimble1998}. 

We thank Anders Mortensen for helpful advice on the manuscript. This work was supported by NSF CUA, and the COMMIT project funded by the ARO. P.F.H. is grateful for the support from the Carlsberg Foundation and the Lundbeck Foundation.

\end{document}